# TiQuant: Software for tissue analysis, quantification and surface reconstruction

Adrian Friebel[1], Johannes Neitsch[1], Tim Johann[1], Seddik Hammad[2,3], Jan G. Hengstler[2], Dirk Drasdo[4,1,#] and Stefan Hoehme[1,*,#]

[1]Interdisciplinary Centre for Bioinformatics (IZBI), University of Leipzig, Germany; [2]Leibniz Research Centre for Working Environment and Human Factors (IfADo), Dortmund, Germany; [3]Department of Forensic Medicine and Veterinary Toxicology, Faculty of Veterinary Medicine, South Valley University, Qena, Egypt; [4]Institut National de Recherche en Informatique et en Automatique (INRIA) Unit Rocquencourt, Le Chesnay Cedex, France

[*]To whom correspondence should be addressed. [#]Contributed equally.

## ABSTRACT

**Motivation:** TiQuant is a modular software tool for efficient quantification of biological tissues based on volume data obtained by biomedical image modalities. It includes a number of versatile image and volume processing chains tailored to the analysis of different tissue types which have been experimentally verified. TiQuant implements a novel method for the reconstruction of three-dimensional surfaces of biological systems, data that often cannot be obtained experimentally but which is of utmost importance for tissue modelling in systems biology.
**Availability:** TiQuant is freely available for non-commercial use at msysbio.com/tiquant. Windows, OSX and Linux are supported.
**Contact:** hoehme@uni-leipzig.de

## 1  INTRODUCTION

During the last decades sophisticated techniques for imaging of cells and tissues have been established. However, the translation of this information into new knowledge is hampered by the difficulty to form consistent hypotheses on the complex interplay between components of biological systems resulting either in physiological function or a diseased state. In recent years mathematical models became increasingly important addressing this question by formalizing the relations between and the interplay of these components in well-controlled model scenarios (Schliess et al, 2014). The construction and parameterization of informative models, however, crucially depends on our ability to quantify structure and dynamic behavior of tissues by image processing and analysis techniques.

Moreover, modeling often requires information that cannot be obtained by available imaging methods either because the structure of interest cannot be accessed experimentally or due to technical limitations. For example, while in most cases the two-dimensional (2D) structure of surfaces in tissues can relatively easily be obtained, the reconstruction of the full three-dimensional (3D) picture usually is much more complicated. Cell margins, for example, can easily be obtained in 2D using a beta-catenin or phalloidin staining, but in 3D available microscopy software typically only allows visualization of cell surfaces while reconstruction and quantification of the corresponding individual cells is not possible. Nevertheless, quantification of individual cell shapes is often required. For example, in liver physiology cell shapes determine cell-cell contact areas that in turn impact metabolic trans-membrane fluxes and thus liver function (Hoehme et al, 2010).

Moreover, surfaces often represent a conceptual or functional boundary rather than an actual biological structure that can be explicitly stained and imaged. For example, liver is subdivided in many small functional units called lobules. Only in few species such as pig the surface of these lobules is represented explicitly by an anatomical membrane-like structure, while in most other species including human no such structure exists. This makes the borders between the functional and anatomical units experimentally very hard to determine. However, detailed knowledge of the full 3D shape of liver lobules is essential to quantify the anatomy of key vascular systems in liver as the sinusoidal or bile networks which is of utmost importance to understand the interplay of components. In this paper, we present comprehensive software for the analysis and quantification of tissue that implements inter alia a novel method for the reconstruction of 3D surfaces. The technique is applicable to well-established and widely used imaging techniques even if staining of some cellular structures is incomplete.

## 2  SOFTWARE

The presented software TiQuant is implemented in portable object-oriented ANSI C++. The GUI is based on QT and supports real-time visualization using OpenGL. TiQuant is embedded in the tissue modelling framework CellSys (Hoehme and Drasdo, 2010) and thus is tightly linked with TiSim, a versatile and efficient simulation environment for tissue models. TiQuant provides an interface to the popular volume visualization tool VolView and further complements its functionality by linking to the open-source libraries ITK and VTK (itk/vtk.org) that implement a wide variety of thoroughly tested state-of-the-art image processing and visualization methods. The image/volume processing chains currently implemented in TiQuant for example include techniques to reconstruct central and portal veins, sinusoidal and bile canaliculi networks as well as hepatic and non-hepatic nuclei from 3D confocal micrographs of liver tissue based on the Adaptive Otsu Thresholding method and a number of morphological operators as described in detail in (Hammad et al, 2014).

## 3  SURFACE RECONSTRUCTION METHOD

Prerequisite of the proposed novel method for the reconstruction of 3D surfaces is the identification of two classes of objects preferentially located 1) in the interior (class *CI*) and 2) on the 3D surface searched for (class *CS*). Naturally, the structures in class *CS* do not need to cover the entire 3D surface. We assume both *CI* and *CS* to be available as binary masks for example obtained by TiQuant's above described segmentation methods. Next, we apply a Signed Maurer Distance Transformation (Maurer et al, 2003) to obtain the Euclidean distance $d_{CI,CS}(\underline{x})$ to the nearest voxel of *CI (CS)* for each point $\underline{x}=(x,y,z)$ in order to compute the gradient function $g(\underline{x})$ using Equ.1:

$$g(\underline{x}) = \beta d_{CI}(\underline{x}) / (\beta d_{CI}(\underline{x}) + (1-\beta) d_{CS}(\underline{x})) \qquad (1)$$

The parameter $\beta$ ($0 \leq \beta \leq 1$) allows balancing the influence of the two classes on the resulting gradient magnitude profile; lower values of $\beta$ reduce the influence of class *CI*. Fig.S1 in the supplement illustrates the behavior of $g(\underline{x})$ for different $\beta$.

In a final step, we apply a variant of the Beucher-Watershed algorithm termed Morphological Watershed (Beare and Lehman, 2006) to $g(\underline{x})$ that aggregates points whose gradient descent leads to the same local minimum. The resulting space partitioning yields sought surfaces in 3D.

In liver, the described method has successfully been used to obtain a) individual cell shapes (Fig.1A-D) and b) the surfaces of liver lobules (Fig.1E). For the reconstruction of cell shapes, cell nuclei were used for class *CI* as they are typically located in the interior of cells while sinusoids and bile canaliculi which are always located at the cell surface were used for class *CO* (Fig.1A).

Since hepatocytes can have more than one nucleus and nuclei don't have to reside in the center of the cell, the influence of *CI* has been decreased to $\beta=0.1$. The result of the surface reconstruction method is shown in Fig.1B/C. We compared the result of our method to a reconstruction based on beta catenin which allowed for precise cell membrane segmentation that we considered a reference cell shape reconstruction. The volume deviation of our method is less than 10% for half the cells and maximally 30% which represents a reasonably accurate reconstruction (see Fig.S2).

In case of liver lobule surface reconstruction, central veins were used for class *CI* as they are per definition located in the center of lobules, while portal veins which are typically located at the border of lobules were used for class *CO*. In this example, both vascular structures were based on µCT images. Since the geometry of central and portal veins equally defines the shape of the lobules, we choose $\beta=0.5$ (Fig.1E). On a modern system (Intel i7 Quadcore), a complete run of the implementation of the cell surface reconstruction method in TiQuant completes in less than 30 minutes for a dataset of size 1024x1024x100 (~100m voxels). The RAM requirement for processing datasets of this size is 16 GB.

## SUMMARY

TiQuant provides a robust and efficient way to reconstruct, visualize and analyze different types of tissue. Additionally, the software implements a novel, widely applicable technique for the reconstruction of surfaces of biological structures based on incomplete information without using explicit staining. The modular and object-oriented implementation permits extensions for new tissue types or new types of analyses that usually can be composed quickly by using existing elementary image processing and analysis modules. New functionalities will be made available in future versions. Open-source release of TiQuant is planned.

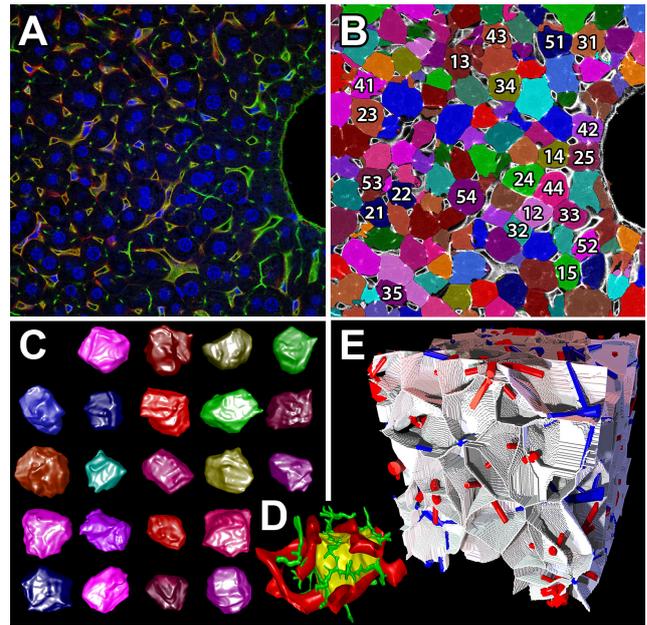

**Fig. 1.** A: Confocal dataset: Nuclei (blue, DAPI), bile canaliculi (green, DPPIV), sinusoids (yellow, DPPIV, DMs). Resulting cell surface reconstruction in (B) 2D and (C) 3D. Marked cells are randomly selected. Numbers in (B) denote row (1st digit) and column (2nd digit) in (C). (D) Single cell in tissue context (yellow: cell, green: bile, red: sinusoids). (E) Reconstruction of lobule surfaces in liver. See Fig.S3 for reconstruction in 2D.

*Funding* by Virtual Liver Network, EU-Notox, BMBF liver simulator, and ANR-iFLOW is gratefully acknowledged.

Supplement

Fig.S1: Gradient function for different parameters $\beta$

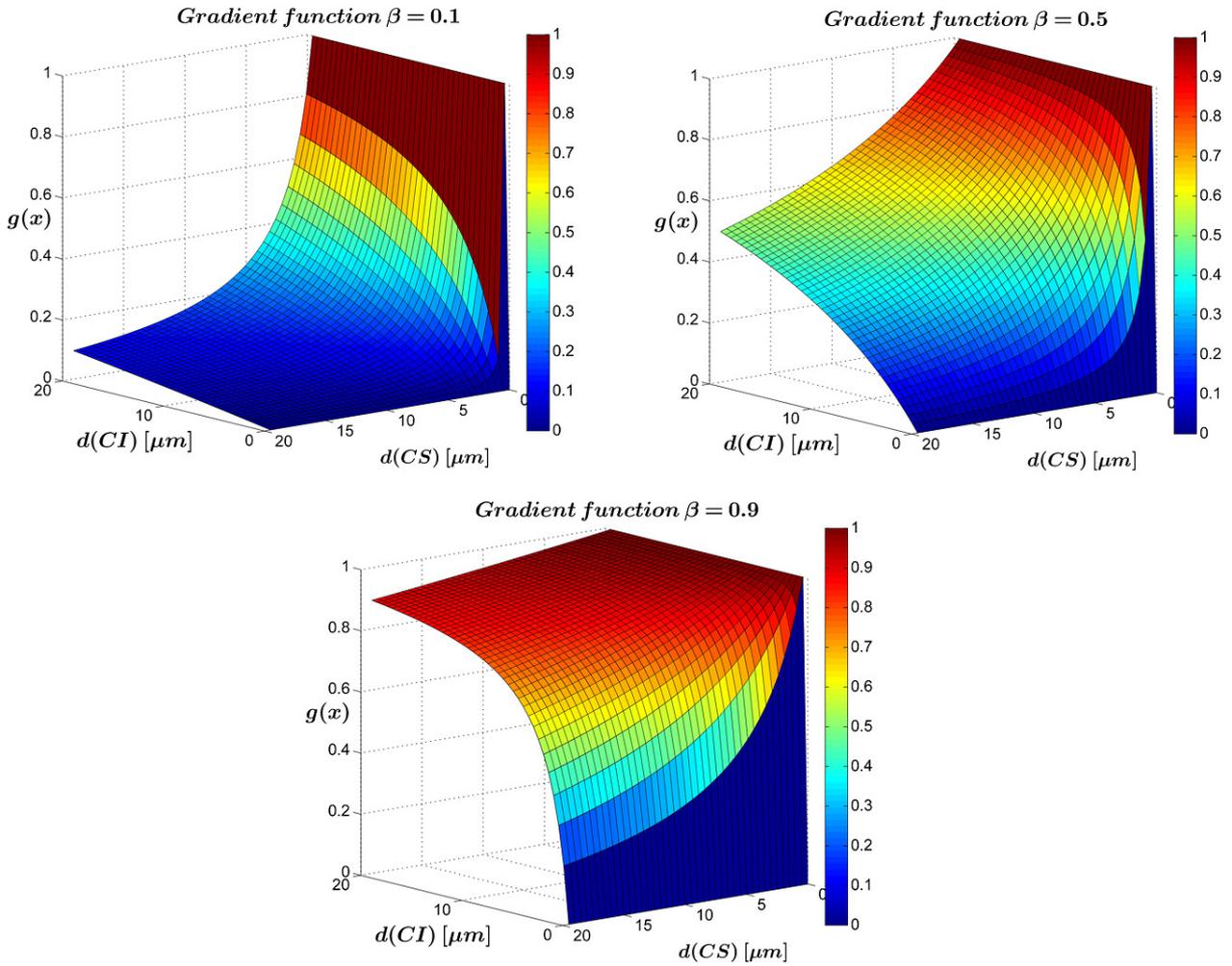

Visualization of gradient $g(x)$ for different $\beta$ that balances the impact of classes CI and CS in the Morphological Watershed algorithm. For $\beta = 0.5$ the impact of CI and CS is equal, the lower $\beta$ the higher the impact of class CS.

Fig.S2: Verification of the new method by comparison with surface reconstruction based on beta catenin staining

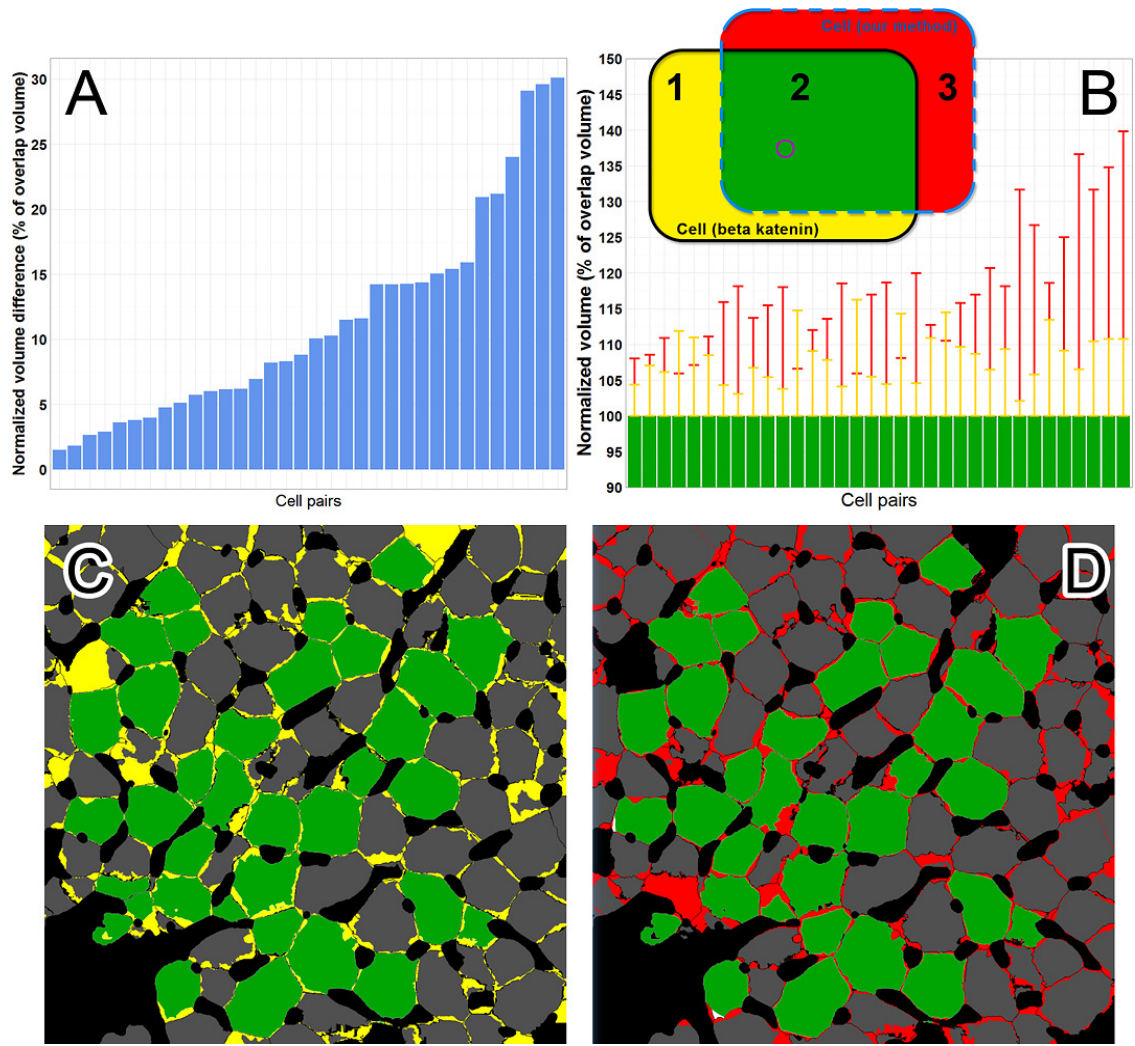

We validate the computational surface reconstruction method described in the main text using a specific marker setup for identification of nuclei (DAPI), bile canaliculi and sinusoids (DPPIV) as well as cell membranes (beta-catenin) in the same specimen. Based on that combination of staining, we compare (1) a reconstruction of cell shapes using the cell membrane marker (beta-catenin) and (2) a second reconstruction using our computational method utilizing nuclei, bile and sinusoids in the same dataset which allows us to directly compare the results of (1) and (2). In the following, cells obtained with (1) are called reference cells and cells obtained with (2) are called approximated cells.

In order to identify overlapping reference / approximated cell pairs, we determine the center of mass (small violet circle in inset of (B)) of a reference cell, and assign the approximated cell at this position as corresponding cell. Thereby, we can identify three spatial classes for cell pairs, as illustrated in the inset of (B): Error-class 1: (1, yellow) is volume that is associated to the cell in the beta-catenin based reference reconstruction but that is not associated to the corresponding cell in our method. Overlap-class: (2, green) represents volume that is associated to the same cell in our method and also in the beta catenin based reconstruction. Error-class 2: (3, red) is volume that is not associated to the cell in the beta-catenin based reconstruction but that is associated to the corresponding cell in our method. (A) illustrates the accuracy of the resulting cell volume. Cell-pairs (each represented by one bar) are numerically sorted and normalized. (B) shows the spatial accuracy of the reconstruction. Again, each bar represents the normalized volume of one cell pair. For each pair the two error bars represent the two error-classes, color coded as in the inset. Both error bars begin at the normalized volume of 100% indicating thereby the volume excess of each error-class in relation to the overlap volume. We only analyzed cells that were not in contact with the data set borders to avoid artifacts. (C) and (D) show a visual representation of a central slice of the data set (similar to Fig.1B in the main text, colors correspond to the inset in (B)). Cells colored in grey are in contact to data set borders. Note, that this comparison includes errors due to possible staining artifacts in the beta catenin staining.

Fig.S3: Lobule shape in 2D

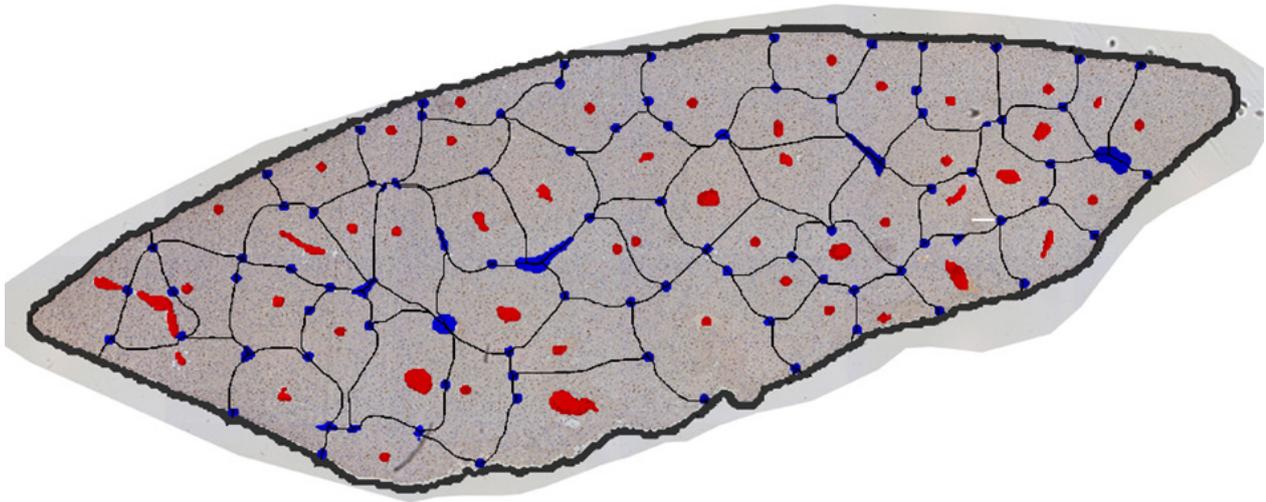

Lobule shape reconstruction by the surface reconstruction method applied to a 2D serial section. Red: central veins, Blue: portal veins, Black: Result of surface reconstruction = lobule boundaries. Image by Rolf Gebhardt (University of Leipzig, Germany).